\documentclass[twoside,12pt]{article}
\usepackage{epsfig}
\usepackage{amsmath}

\newcommand{\be}{\begin{equation}}
\newcommand{\ee}{\end{equation}}
\newcommand{\bea}{\begin{eqnarray}}
\newcommand{\eea}{\end{eqnarray}}

\date{\small $^*$ Talk given by B.\ G.\ at the Erice summer school on Nuclear
Physics, Sept.\ 16 -- 24, 2003, Erice, Italy}
\begin{document}

\title{ \vspace{5mm} QCD Propagators at non-vanishing temperatures$^*$}
\author{B.\ Gr{\"u}ter$^1$, R.\ Alkofer$^1$, A.\ Maas$^2$, J.\ Wambach$^2$\\
{$^1$ Inst.\ f.\ Theoretische Physik, Universit{\"a}t T{\"u}bingen, D-72076 T{\"u}bingen, Germany} \\
{$^2$ Inst. f{\"u}r Kernphysik, TU Darmstadt, D-64289 Darmstadt, Germany}}
\maketitle
\begin{abstract} 
We investigate the behaviour of the gluon and ghost propagators, especially their infrared properties, 
at non-vanishing temperatures. To this end we solve their Dyson-Schwinger equations on a
torus and find an infrared enhanced ghost propagator and an infrared vanishing
gluon propagator.
\end{abstract}

Lattice simulations provide evidence that zero-flavour QCD undergoes a 
"deconfining" phase transition at some critical temperature.
The properties of the gluon and the ghost propagator at finite temperature 
can provide insights into the mechanisms of this phase transition 
and they are related to the issue of confinement and deconfinement.

We study the Dyson-Schwinger equations (DSEs) of these propagators in Landau gauge for which
a well understood truncation scheme at zero temperature is available \cite{Fischer:2003rp}.
The infrared behaviour of the gluon and ghost propagators can be determined analytically 
and we obtain power-laws.
This is consistent with the Kugo-Ojima confinement picture for Landau gauge, which requires 
the ghost propagator to diverge stronger than a free propagator at
zero momentum. On the other hand the gluon propagator is infrared suppressed
($D_{\mu \nu}(k\rightarrow 0)=0$). Thus it violates positivity 
and the gluons are removed from the physical spectrum \cite{Alkofer:2003jk}.
Since this infrared singular structure is expected to persist for non-vanishing 
temperatures a continuum based method for studying the deconfinement phase 
transition is desirable.

At non-vanishing temperatures the  analytical determination of the infrared behaviour 
has not been achieved yet, thus we need an infrared regulator. A possible one is to compactify space-time to
a four dimensional torus by imposing periodic boundary conditions on the fields.
This has been done already at zero temperature \cite{Fischer:2002eq}.

\begin{figure}
\begin{minipage}[b]{9.2 cm}
\begin{center}
 \epsfig{file=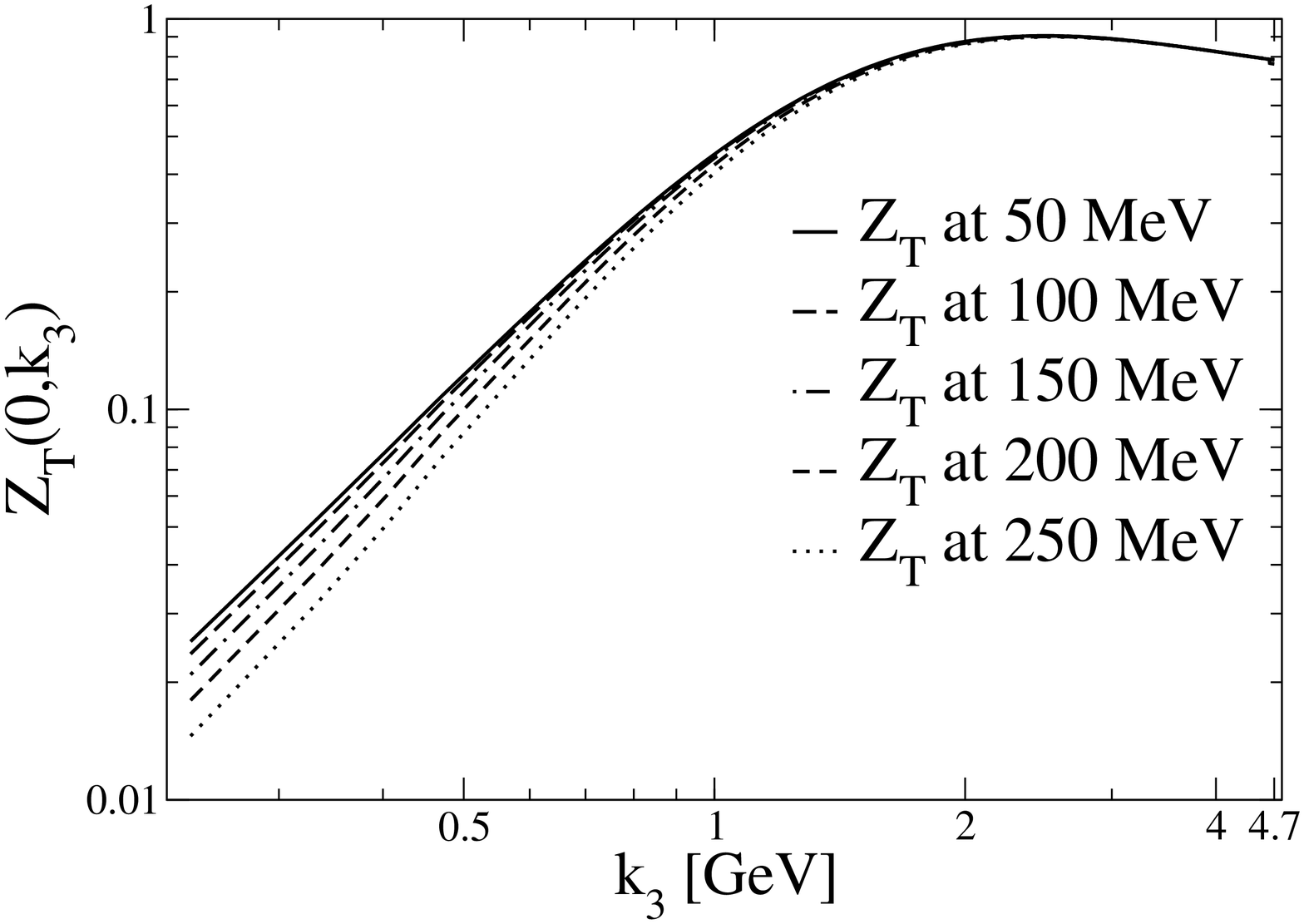,scale=0.3}   
 \parbox{8 cm}{
 \caption{$Z_T$ (from $20^3$ momentum grid)} 
  \label{fig:ZT_32}
 }
 \end{center}
\end{minipage}
\begin{minipage}[b]{9.2 cm}
\begin{center}
  \epsfig{file=ZL_20.eps,scale=0.3}
  \parbox{8 cm}{
 \caption{$Z_L$ (from $20^3$ momentum grid)}
 \label{fig:ZL_32}}
 \end{center}
\end{minipage}
\end{figure}

The truncation scheme at zero temperature described in \cite{Fischer:2002eq}
is extended to finite temperatures in the Matsubara-Formalism (imaginary time formalism) for
thermodynamical equilibrium. The DSEs in pure Yang-Mills-Theory for the gluon and the ghost
propagator are considered. 
Since the heat bath defines a prefered reference frame,
the full gluon-propagator in Landau gauge now depends on two independent transverse
tensor structures, a heat-bath transverse and longitudinal part, leading to the following 
decomposition \cite{Kap93}: 
\begin{align}
\label{Dmunu}
  D^{ab}_{\mu \nu}(k)&:=\frac{\delta^{ab}}{k^2}\left(P_{T\, \mu \nu}(k) Z_T(k_0,|\vec k|) + 
  P_{L\, \mu \nu}(k) Z_L(k_0,|\vec k|)\right), \quad P_{T\, i j}(k)=\delta_{ij} -\frac{k_i k_j}{\vec k^2}, \\
  P_{T\, 0 0}&=P_{T\, i 0}=P_{T\, 0 i}=0, \quad 
  P_{L\, \mu \nu}(k) =P_{\mu \nu}(k)-P_{T\, \mu \nu}(k), \quad P_{\mu \nu}=\delta_{\mu \nu}-\frac{k_{\mu} k_{\nu}}{k^2} 
\end{align}
with $i,j=1,2,3;\; \mu,\nu=1..4$ and $k^2=k_0^2+\vec k^2$. 
The ghost propagator being a Lorentz scalar reads: $D^{ab}_G(k):=-\frac{\delta^{ab}}{k^2}G(k_0,|\vec k|)$.  
Note, that the dressing functions of the gluon ($Z_T$ and $Z_L$) and the ghost ($G$)
propagators depend on the Matsubara-frequency $k_0=2\pi n T \, (n=..-2,-1,0,1,2,..)$ and the three momentum
$\vert \vec k \vert$ separately. So the DSEs for the propagators lead to a coupled system of 
three coupled integral equations for three scalar functions \cite{Maas:2002}. In compactified space-time 
non-vanishing temperatures render the torus asymmetric with a "shorter" time direction:
\vspace{-0.1cm}
\begin{align}
T \sum_{n=-\infty}^{+\infty} \int \frac{d^3 q}{(2 \pi)^3} \rightarrow \frac{T}{L^3} \sum_{n=-\infty}^{+\infty} \sum_{j_1,j_2,j_3}
\end{align}
\vspace{-0.1cm}
with $L_0:=\beta=1/T$ and $L_0\ll L$.
The numerical results from an asymmetric torus for the dressing functions at small temperatures 
indicate a rather weak temperature dependence. With increasing temperature the heat-bath transverse dressing
$Z_T$ of the gluon gets suppressed (fig.\ \ref{fig:ZT_32}) while the longitudinal part $Z_L$ gets enhanced
(fig.\ \ref{fig:ZL_32}). However, the power law resists, especially for the ghost propagator.

In the $T\rightarrow \infty$ limit the four dimensional Yang-Mills theory is reduced to a three dimensional one with
an additional Higgs field. 
Starting from the respective DSEs \cite{Maas:2002} the infrared analysis for the system of gluon, ghost and
"Higgs" has been performed and first numerical results have been obtained. The full "Higgs" propagator is close
to a massive tree-level propagator indicating the existence of a chromoelectric screening mass at high temperatures.
However, in the chromomagnetic sector an overscreening of soft modes occurs and the gluon propagator vanishes at
zero momentum.

The further aim of these studies is to provide a dynamical description of the expected phase transition 
from the two different regions of the phase diagram by use of the methods sketched here.
In addition the studies at low temperatures will be supplemented by more complete and accurate numerical solutions
of the DSEs. Thereby we aim at a dynamical description of the physical excitations across the phase transition.

\end{document}